\documentclass[twocolumn]{article}
\usepackage[utf8]{inputenc}
\usepackage[english]{babel}
\usepackage[T1]{fontenc}
\usepackage{amsmath}

\usepackage{tikz}
\usepackage{lipsum}
\usepackage{doi}
\usepackage{textcomp}
\usepackage{amssymb}
\usepackage{graphicx}
\usepackage{braket}
\usepackage{url} 
\usepackage{comment} 
\usepackage{subfigure}
\usepackage[makeroom]{cancel}
\usepackage{authblk}

\begin{document}

\title{Separating a particle's mass from its momentum}

\author[a,*]{Mordecai Waegell}
\author[a,b]{Jeff Tollaksen} 
\author[a,b,c,d]{Yakir Aharonov}

\affil[a]{\small{Institute for Quantum Studies, Chapman University, 1 University Dr., Orange, CA 92866, USA}}
\affil[b]{Schmid College of Science and Technology, Chapman University, 450 N Center St., Orange, CA 92866, USA}
\affil[c]{Iyar, The Israeli Institute for Advanced Research, POB 651 Zichron Ya'akov 3095303, Israel}
\affil[d]{School of Physics and Astronomy, Tel Aviv University, Tel Aviv, Israel}
\affil[*]{Corresponding Author, email: waegell@chapman.edu}



\maketitle

\begin{abstract}
The Quantum Cheshire Cat experiment showed that when weak measurements are performed on pre- and post-selected system, the counterintuitive result has been obtained that a neutron is measured to be in one place without its spin, and its spin is measured to be in another place without the neutron.  A generalization of this effect is presented with a massive particle whose mass is measured to be in one place with no momentum, while the momentum is measured to be in another place without the mass.  The new result applies to any massive particle, independent of its spin or charge.  A \textit{gedanken} experiment which illustrates this effect is presented using a nested pair of Mach-Zehnder interferometers, but with some of the mirrors and beam splitters moving relative to the laboratory frame.  The titular interpretation of this experiment is extremely controversial, and rests on several assumptions, which are discussed in detail.  An alternative interpretation using the counterparticle model of Aharonov et al. is also discussed.
\end{abstract}

\section{Introduction}

The time-symmetric quantum formalism of Aharonov, Bergmann, and Lebowitz \cite{aharonov1964time} offers a unique perspective on situations where measurements are conditioned on a particular post-selection of the measured system, and if the coupling of the measurement device and system is so weak that it reveals very little information per run, the conditioned ensemble average, called a weak value \cite{aharonov1988result,dressel2015weak}, can exhibit counterintuitive behavior.  Thinking about experiments in terms of time-symmetric boundary conditions, i.e, a pre-selected state, a post-selected state, and the weak values which are defined by this pair during the intermediate time, has led to a number of interesting results that would have been much less obvious in the standard treatment \cite{aharonov2013quantum, vaidman2013past,  aharonov2015current, aharonov2016quantum, waegell2017confined, aharonov2018weak, waegell2018contextuality, waegell2023quantum}.  To apply this reasoning, one must think of both the pre-selected (prepared) and post-selected (outcome) state as existing `first,' and the choice of intermediate measurement settings as existing `second,' which is quite contrary to our usual intuition that `first' and `second' should match their temporal order.

Our goal in this article is to present a variation of the Quantum Cheshire Cat experiment \cite{aharonov2013quantum,denkmayr2014observation}, where the mass and spin of a single neutron were separated in the sense that post-selected weak measurements of the position of the neutron find it in location 1 and not location 2, while weak measurements of the neutron spin find spin in location 2 and not in location 1.  Related experiments have also been performed with photons \cite{sahoo2023unambiguous}, and it has inspired a number of interesting ideas \cite{das2020can, liu2020experimental, pan2020disembodiment, chowdhury2021wave, aharonov2021dynamical, kim2021observing, ghoshal2023isolating, wagner2023quantum,  hance2024dynamical,aharonov2024angular}. 

Already in the above paragraph, an important interpretive assumption has been used, which is that the weak measurements can actually \textit{find} anything about any individual neutron or its spin, since the weak values are only obtained as averages over many experiments with the same pre- and post-selection.  This is unlike an eigenvalue which can be observed on each individual run using standard strong (projective) measurements.  For a strong measurement of a projection operator, the outcome eigenvalue 1 means you have found the system in the corresponding state, and eigenvalue 0 means you have found it in an orthogonal state.  It is a hotly contested assumption \cite{vaidman2013past, li2013comment, vaidman2013reply,correa2015quantum, duprey2017null,  sokolovski2018comment, duprey2018reply, englert2017past,peleg2019comment,englert2019reply, matzkin2019weak, bhati2022weak, hance2023weak, vaidman2023comment,hance2023reply} that measuring the weak value of a projector to be 1 corresponds to finding every element in the measured ensemble to actually have the physical properties associated with that state, or for a weak value of 0, to actually have the physical properties associated with an orthogonal state.  This assumption is motivated in several ways.  First, there is the ABL rule for dichotomic observables \cite{waegell2023quantum}, which says that, for a given pre- and post-selection, if an intermediate strong measurement had been performed instead of a weak one, and we would have gotten a certain projector to have eigenvalue 1 (0) with certainty for every element, then the corresponding weak value 1 (0) that appears when no intermediate strong measurement is performed should be interpreted as a true property, for each individual element of the ensemble.  Second, on each run of the experiment, the wavefunction of the measurement device pointer, conditioned on the pre- and post-selection of the measured system, contains the weak value as a parameter, and this parameter is uniquely determined by the pre- and post-selected states of the measured system.  By the PBR theorem  \cite{pusey2012reality} this pointer wavefunction is ontologically distinct from any other wavefunction, with any other value of the parameter, and thus the weak value is a distinct property of the individual pre- and post-selected elements, and not just something that appears in the ensemble average.  Third, the weak values can be empirically observed for individual pre- and post-seleted systems using a feedback compensation protocol \cite{hofmann2021direct}.  There are other ways to argue that the weak values are true physical properties of individual pre- and post-selected systems, and these arguments are also generally contested.   

Regardless of assumptions, the weak values are an observable prediction of quantum mechanics, so they must represent some physical property of the ensemble of identically pre- and post-selected systems. What is true in these experiments is that the weak values obtained by weak measurements show the Quantum Cheshire Cat (QCC) effect, but only if they are interpreted as describing properties of the system in the same way that eigenvalues are interpreted.  The same measured weak values can be given other interpretations \cite{saeed2021quantum}, which will be discussed elsewhere in this article, but ultimately, we think these counterintuitive experimental results are of great interest, and begging for interpretation, and that nothing highlights this better than the controversial QCC interpretation.

In the present article, we consider the more general case of a massive particle whose position can be detected via its gravitational field, and whose momentum can be detected via an impulsive coupling, and show that post-selected gravitational weak measurements of the position of the mass find it in location 1 and not location 2, while impulsive weak measurements of the momentum find nonzero momentum in location 2 and zero in location 1.  In this sense, we show that the momentum of the particle has been separated from its mass, using the same basic assumptions as in the original QCC.  Most of the kinetic energy is separated in the same way, although an arbitrarily small amount remains with the mass.  This effect is related to interaction-free energy transfer \cite{elouard2020interaction}, where the weak value of the projector onto one arm of the interferometer is zero, while the weak value of the energy on that arm is nonzero.

Our results apply to any massive particle, and thus generalize the original Quantum Cheshire Cat Effect.  In the \textit{gedanken} experiment we introduce, the position of the mass is detected via the gravitational force, but the mirrors, beam splitters, and impulsive detectors must interact with the particle through other forces, regardless of which forces they are.

The remainder of this article is organized as follows.  We begin by introducing the pre- and post-selected states for the new effect in Sec. \ref{sec:PPS}.  Then the relevant weak values are derived and their interpretation is discussed in Sec. \ref{sec:weak}.  Next, some details of how one might perform a gravitational weak measurement are given in Sec. \ref{sec:WM}.  We then give the \textit{gedanken} experiment to test the effect in Sec. \ref{sec:Gedanken}, and some analysis of the weak values in different parts of the experiment in Sec. \ref{sec:analysis}, where the connection to quantum contextuality is discussed.  In Sec. \ref{sec:counter} we give the interpretation of this experiment using the counterparticle model of Aharonov et al., which is fundamentally different than the QCC interpretation, and is free from logical paradoxes.  We end with a few concluding remarks in Sec. \ref{sec:conc}.

\section{The Pre-Selected and Post-Selected States}\label{sec:PPS}

We begin by constructing the desired pre-and post-selected states to show the effect, and then we provide some details of a \textit{gedanken} experiment to produce these states.  We begin by constructing three different normalized Gaussian states, $f_\pm(x)$ and $g(x)$.
The functions $f_\pm(x)$ are centered at the origin with average momenta given by $k_0 \pm k_1$, while the function $g(x)$ has zero average momentum and it is centered at $x_0$.
\begin{equation}
    {f}_\pm(x) = \Big(\frac{2\Delta k_f^2}{\pi}\Big)^{1/4} e^{i(k_0 \pm k_1)x} e^{-x^2\Delta k_f^2}
\end{equation}
\begin{equation}
    \tilde{f}_\pm(k) = \Big(\frac{1}{2\pi\Delta k_f^2}\Big)^{1/4}e^{-(k - (k_0 \pm k_1 ))^2/4\Delta k_f^2}
\end{equation}
\begin{equation}
    {g}(x) = \Big(\frac{2\Delta k_g^2}{\pi}\Big)^{1/4} e^{-(x-x_0)^2\Delta k_g^2}
\end{equation}
\begin{equation}
    \tilde{g}(k) = \Big(\frac{1}{2\pi\Delta k_g^2}\Big)^{1/4}e^{ikx_0}e^{-k^2/4\Delta k_g^2}
\end{equation}

From $f_\pm$ we construct two new (unnormalized) states $h_\pm$ as,
\begin{equation}\begin{array}{cc}
      h_+ \equiv f_+ + f_-,  & h_- \equiv f_+ - f_-.
\end{array}
\end{equation}
Note that $f_+$ and $f_-$ are not orthogonal, but $h_+$ and $h_-$ are,
\begin{equation}
\begin{array}{cc}
     \langle f_+ | f_- \rangle = e^{-k_1^2 / 2 \Delta k_f^2},   &    \langle h_+ | h_- \rangle = 0.
\end{array}
\end{equation}
From these and $g$ we construct our (unnormalized) pre- and post-selected states, $\psi$ and $\phi$, respectively.
\begin{equation}
\begin{array}{cc}
      \psi =  g +h_+, &   \phi =  g +h_-.
\end{array}
\end{equation}

\section{Weak Values} \label{sec:weak}

The weak value of any operator $\hat{A}$ on this system is defined as
\begin{equation}
    \hat{A}_w \equiv \frac{\langle \phi |\hat{A}|\psi\rangle}{\langle\phi|\psi\rangle}.  \label{weak}
\end{equation}
We set $x_0 \gg \Delta x_g = 1/(2 \Delta k_g)$ and $x_0 \gg \Delta x_f = 1/(2 \Delta k_f)$   so that $\langle f_\pm | g \rangle \approx 0$ and thus $\langle h_\pm | g \rangle \approx 0$.  Then we have
\begin{equation}
    \langle \phi| \psi\rangle \approx \langle g|g\rangle = 1,
\end{equation}
so we can neglect the denominator in Eq. \ref{weak} when finding weak values.

We now define projectors onto two separate regions of the $x$-axis; $\hat{\Pi}_h$ is the region where $h_\pm(x)$ have support and $g(x)$ has almost none, and vice versa for $\hat{\Pi}_g$.  As an approximation, we will define these projectors so that they perfectly isolate the two functions.  Applying these operators to $\psi$ and $\phi$ we have,
\begin{equation}
    \begin{array}{cc}
       \hat{\Pi}_h |\psi\rangle = |h_+\rangle,  & \hat{\Pi}_g |\psi\rangle = |g\rangle, \\\\
       \langle \phi |\hat{\Pi}_h^\dag  = \langle h_-|,  & \langle \phi | \hat{\Pi}_g^\dag  = \langle g|.
    \end{array}
\end{equation}

We are now in a position to consider the weak values of the projectors onto these regions, along with the localized energy and momentum in each region.

For the projectors alone we have,
\begin{equation}
\begin{array}{c}
        (\hat{\Pi}_h^\dag\hat{\Pi}_h)_w =\langle \phi|\hat{\Pi}_h^\dag \hat{\Pi}_h |\psi\rangle = 0,  \\\\  (\hat{\Pi}_g^\dag\hat{\Pi}_g)_w = \langle \phi|\hat{\Pi}_g^\dag \hat{\Pi}_g |\psi\rangle = 1,
\end{array}
\end{equation}
so the particle seems be located entirely in the region spanned by $\hat{\Pi}_g$.

For the localized momentum we have,
\begin{equation}
\begin{array}{c}
       (\hat{\Pi}_h^\dag \hat{p}\hat{\Pi}_h)_w =  \langle \phi| \hat{\Pi}_h^\dag \hat{p}\hat{\Pi}_h |\psi\rangle = 2p_1 = 2\hbar k_1, \\\\ (\hat{\Pi}_g^\dag \hat{p}\hat{\Pi}_g)_w = \langle \phi|\hat{\Pi}_g^\dag \hat{p}\hat{\Pi}_g |\psi\rangle = 0,
\end{array}
\end{equation}
so the localized momentum at the apparent location of the particle ($\hat{\Pi}_g$) is zero, while the localized momentum in the region where the particle is not located ($\hat{\Pi}_h$) is $2p_1$.

For the localized energy we have,
\begin{equation}
\begin{array}{c}
        (\frac{\hat{\Pi}_h^\dag \hat{p}^2\hat{\Pi}_h}{2m})_w = \frac{1}{2m}\langle \phi|\hat{\Pi}_h^\dag \hat{p}^2\hat{\Pi}_h |\psi\rangle =  \frac{2p_0p_1 }{m}= \frac{2\hbar^2k_0k_1}{m}
       ,  \\\\  (\frac{\hat{\Pi}_g^\dag \hat{p}^2\hat{\Pi}_g}{2m})_w = \frac{1}{2m}\langle \phi|\hat{\Pi}_g^\dag \hat{p}^2\hat{\Pi}_g |\psi\rangle = \frac{ \Delta p_g^2}{2m} =  \frac{\hbar^2 \Delta k_g^2}{2m}.
\end{array}
\end{equation}
Now, in the limit that $\Delta k_g = (1/2\Delta x_g)$ is very close to zero, the localized energy at the apparent location of the particle ($\hat{\Pi}_g$) is also nearly zero, while the localized energy in the region where the particle is not located ($\hat{\Pi}_h$) may be much larger.

So for these pre- and post-selected states, with parameters chosen so that $0 \ll \Delta x_g \ll x_0$ and $\Delta x_f \ll x_0$, we have our generalization of the Quantum Cheshire Cat effect: The mass is located in one place without momentum, and the momentum is located elsewhere without mass.  Note that neither $k_0$ nor $k_1$ needed to be constrained to see this effect, so there is no bound on the energy or momentum of the particle.

Now, we must acknowledge that we are making several more contentious assumptions.

First, we assume that the position of the active gravitational mass of the particle corresponds to the weak value of the projector onto position $\vec{x}$, so that we can weakly measure these projectors using the gravitational interaction.  We have already discussed why we interpret the weak value as a physical property, and given that gravity is a distance-based force, it seems that no other operator makes sense.  We offer no other defense of this assumption.

Second, we assumed that operators like $\hat{\Pi}_h^\dag \hat{p}\hat{\Pi}_h$, $\hat{\Pi}_g^\dag \hat{p}\hat{\Pi}_g$, $\frac{1}{2m}\hat{\Pi}_h^\dag \hat{p}^2\hat{\Pi}_h$  and $\frac{1}{2m}\hat{\Pi}_g^\dag \hat{p}^2\hat{\Pi}_g$ represent the local momentum and local energy in the regions $h$ and $g$.  Certainly the Heisenberg uncertainty relation forbids us from finding the exact value of momentum in a finite region, but the measured weak values are ensemble average like expectation values, so we're not claiming to know the exact momentum case by case.  Furthermore, the width of the regions $h$ and $g$ can be made large enough, so that the restriction on momentum uncertainty is negligible.  More importantly, these operators correspond to real operations we can perform in the lab.  For example, if we measure the momentum on one arm of the experiment using a device that is insensitive to momentum anywhere else, then this device can be treated as only measuring momentum on that arm.  The same is true for the energy.  For this scenario, it would not make sense to use observable operators that are sensitive to what is happening far away from the detector, and incorporating projectors that clip off the wavefunctions in such regions is arguably a reasonable approach.  It is easy to see that the eigenstates of these operators are the clipped versions of the standard momentum and energy eigenstates, which only span that region of space, so these observables represent truly localized properties in that region.  The assumption that experimental procedures which are physically confined to one spatial region are `local operations' in that region is standard in many applications, so we won't defend it further here.

That said, we acknowledge that we have made these interpretive assumptions, and that other interpretations are possible.

\section{Weak Measurement}\label{sec:WM}

Impulsive weak measurement has been explained in detail in the literature, so we won't repeat it here.  The gravitational measurement of projectors onto position is somewhat new, so we explain one possible protocol here.  In this protocol, a pointer particle is sent into a parallel trajectory with the probe particle, very close together, and interacting by only gravity, such that two proceed side-by-side.  Then, if the probe particle is present, the gravitational attraction will slightly deflect the pointer, and if the probe is absent, the pointer will be unaffected.  This can be arranged so that the maximum deflection of the pointer is much less than the width of its wave packet, meaning that measuring the pointer position constitutes a weak measurement of the probe position, where the outcome can only be resolved for a large ensemble.

What follows is an idealized and simplified version of the gravitational weak measurement process.  We will approximate each particle as a point located at its position expectation value, and use the classical calculation of gravitation acceleration and deflection.  We will treat the probe particle as having only two states relative to the gravitational detector -- present and absent -- where present will correspond to some particular arrangement that leaves the pointer completely unchanged if the probe is absent, and leaves it deflected by distance $d$ if the probe is present.  We neglect the momentum of the deflected pointer, as well as the deflection of the probe.

The normalized initial state of the probe and pointer is,
\begin{equation}
    |\psi\rangle G(x) = \big(a|\textrm{present}\rangle +b |\textrm{absent}\rangle\big) G(x),
\end{equation}
where $ |\psi\rangle = \big(a|\textrm{present}\rangle +b |\textrm{absent}\rangle$ is the pre-selected probe state, and $G(x)$ is the normalized Gaussian pointer wavefunction with width $w \gg d$.
After the gravitational interaction is completed, we have the weakly entangled state,
\begin{equation}
    a|\textrm{present}\rangle G(x-d) +b |\textrm{absent}\rangle G(x).
\end{equation}
If we post-select the probe state as $|\phi\rangle = \big(c|\textrm{present}\rangle +d |\textrm{absent}\rangle $ and renormalize, then the pointer state is approximately,
\begin{equation}
    \frac{ac^*G(x-d) + bd^*G(x)}{ac^* + bd^*} \approx G\big(x - (|\textrm{present}\rangle\langle \textrm{present}|)_w\big),
\end{equation}
where the last step is the standard weak value derivation, done using the first order approximation of the exponential.

The weak value can only be resolved by measuring a large ensemble of pointer states prepared in this way, but as discussed above, the PBR theorem tells us there is no ontological overlap between pointer states with different weak values, and it is the pre- and post-selected states of the probe that gives the pointer this value, so the weak value represents a physical property of each individual probe particle.

\section{\textit{Gedanken} Experiment} \label{sec:Gedanken}

To measure the momentum and energy, it should be sufficient to perform a local impulsive coupling with a broadly spread pointer in region $h$ and another in region $g$.  The pointers will weakly measure that there is momentum in region $h$ and none in region $g$.  It does not matter what force mediates the impulsive couplings, mirrors, and beam splitters.  Another pair of pointers will couple via the gravitational interaction, and these will find mass located in region $g$ but none in region $h$.

\begin{figure}[t!]
    \centering
    \includegraphics[width= 3in]{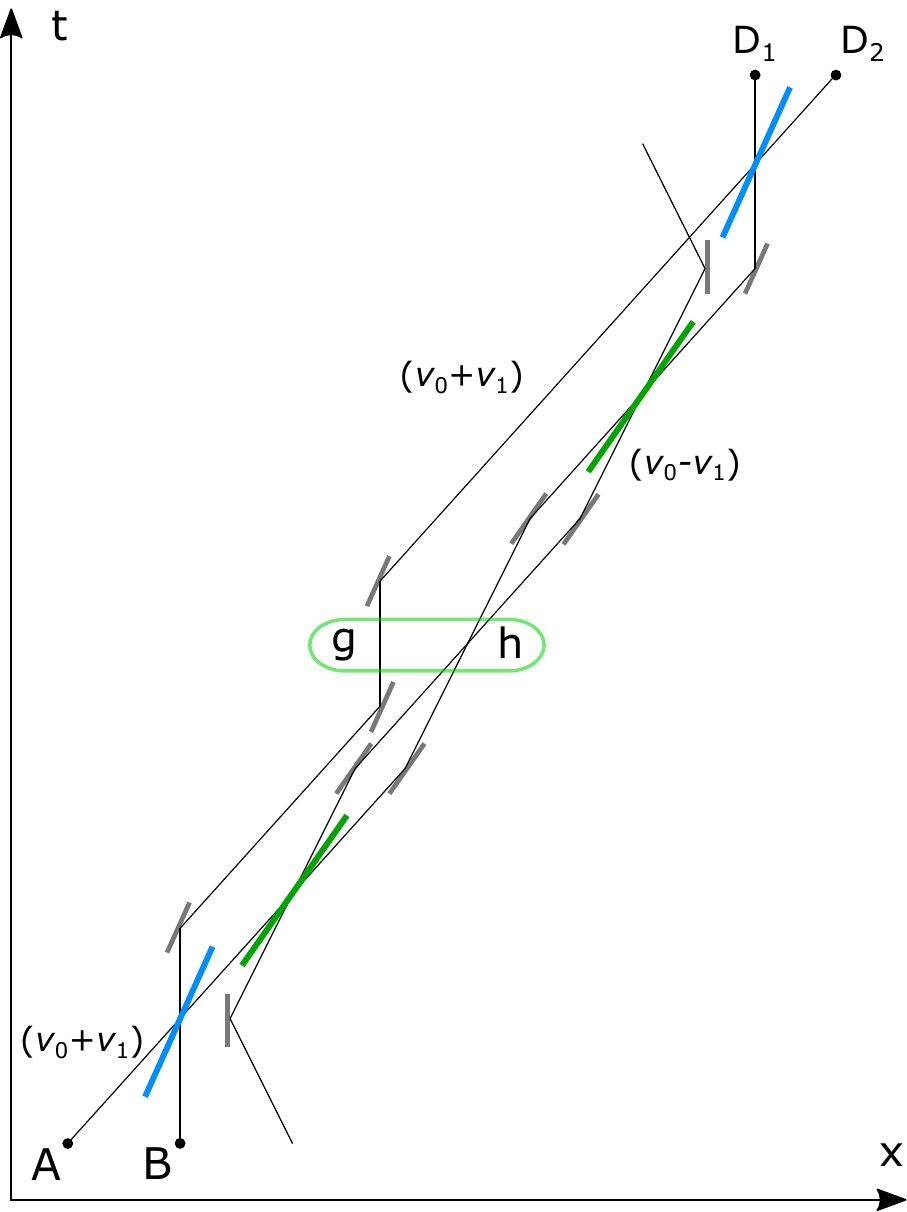}
    \caption{The one-dimensional experimental setup for the energy-momentum Quantum Cheshire Cat effect, with all forward and backward paths shown.  The tilts of the mirrors (gray), beam splitters (blue and green), and trajectories show their velocities ($\frac{1}{2}(v_0+v_1)$ for blue, and $v_0$ for green), while the sources and detectors are at the shown events.  The mirrors and beam splitters obviously exist at all times, but to avoid clutter, they are only shown at events where they affect the beam.}
    \label{PPS}
\end{figure}

To accomplish this, we consider a setup which has some resemblance to the nested interferometers used by Vaidman \cite{vaidman2013past}, but we will use a massive particle moving at nonrelativistic speeds, and some of the mirrors and beam splitters that are used to redirect the particle will also be moving.  The experimental setup is shown in Fig. \ref{PPS} as a space-time diagram depicting motion in one dimension.  Because this is one-dimensional, the tilt of the mirrors and beam splitters represents their velocities (it shows the direction of their world-lines), and likewise for the tilt of the particle trajectories.

To prepare the desired pre-selected state, we begin by sending a Gaussian wave packet $f_+$ ($v = \hbar k /m$ and $v_+ \equiv v_0 + v_1$) into our apparatus, and then passing it through a beam splitter moving with velocity $\frac{1}{2}v_+$, such that one path comes to rest at $x_0$ ($g$) and the other propagates with velocity $v_+$ ($f_+$, and we have $\Delta k_f = \Delta k_g$, since they split off from the same packet).   To see this, consider the situation in the rest frame of the beam splitter, where the velocity of the incoming particle is $v_+' = v_+ - \frac{1}{2}v_+ = \frac{1}{2}v_+$.  The resting beam splitter then naturally results in a reflected path with velocity $-\frac{1}{2}v_+$, and a transmitted path with velocity $\frac{1}{2}v_+$.  Returning to the original frame where the beam splitter has velocity $\frac{1}{2}v_+$, we see that the reflected path has velocity $v_r = -\frac{1}{2}v_+ + \frac{1}{2}v_+ = 0$ and the transmitted path has the original velocity $v_t = \frac{1}{2}v_+ + \frac{1}{2}v_+  = v_+$.  Also, reflection from mirrors moving at $\frac{1}{2}v_+$ swaps between $v_+$ and $0$.

Similarly, when a particle with velocity $v_+$ is incident on another beam splitter with velocity $v_0$, the reflected path has velocity $v_r = v_- \equiv (v_0 - v_1)$, and $v_+$ for the transmitted path.  Also, reflection from mirrors moving at $v_0$ swaps between $v_+$ and $v_-$.

\begin{figure}[t!]
    \centering
    \includegraphics[width= 3in]{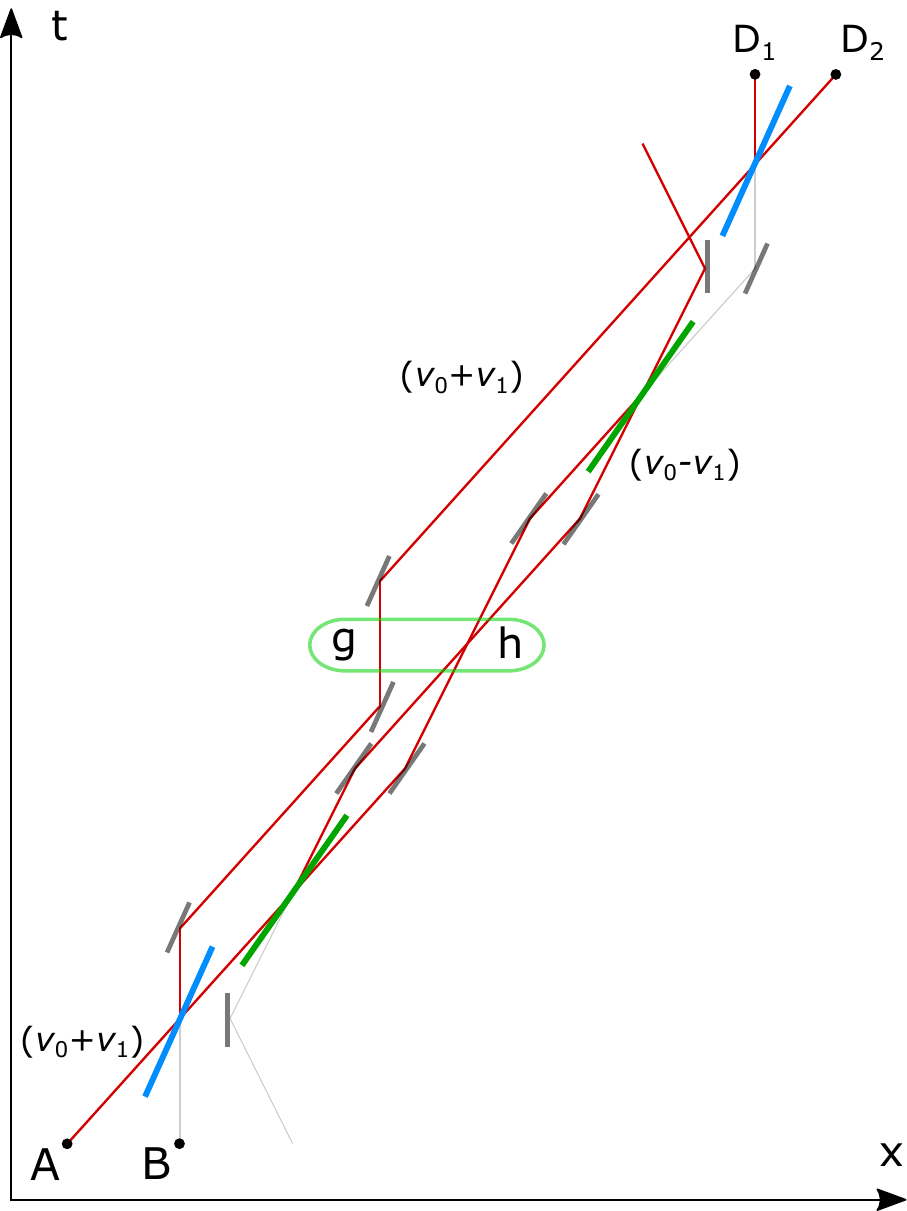}
    \caption{The experimental setup with the particle pre-selected to enter from source A with velocity $v_+ = v_0 + v_1$.  All subsequent places where the pulse has nonzero support are shown in red.}
    \label{PPS_Pre}
\end{figure}

\begin{figure}[t!]
    \centering
    \includegraphics[width= 3in]{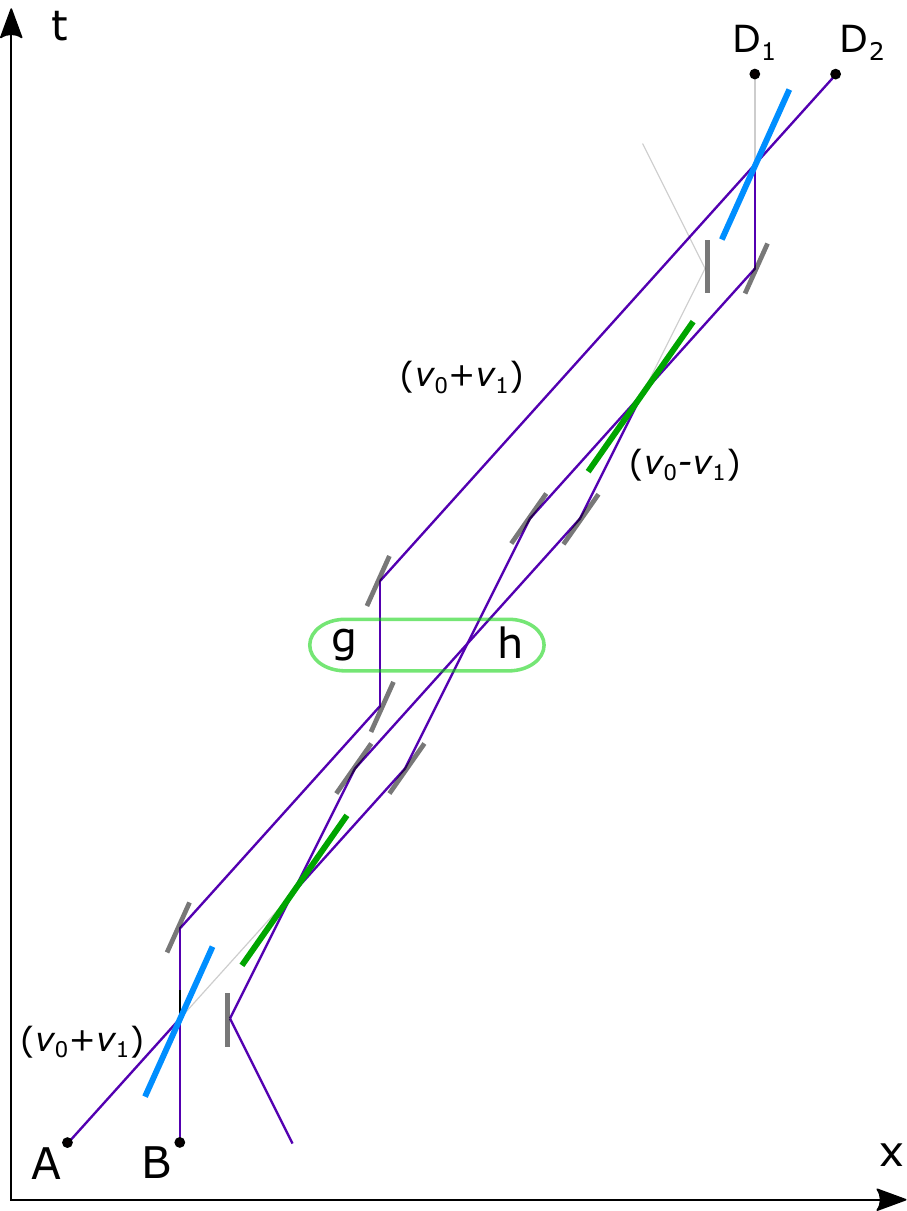}
    \caption{The experimental setup with the particle post-selected to exit at detector D$_2$ with velocity $v_+ = v_0 + v_1$.  All places where the back-propagated pulse has nonzero support are shown in dark purple.}
    \label{PPS_Post}
\end{figure}

\begin{figure}[t!]
    \centering
    \includegraphics[width= 3in]{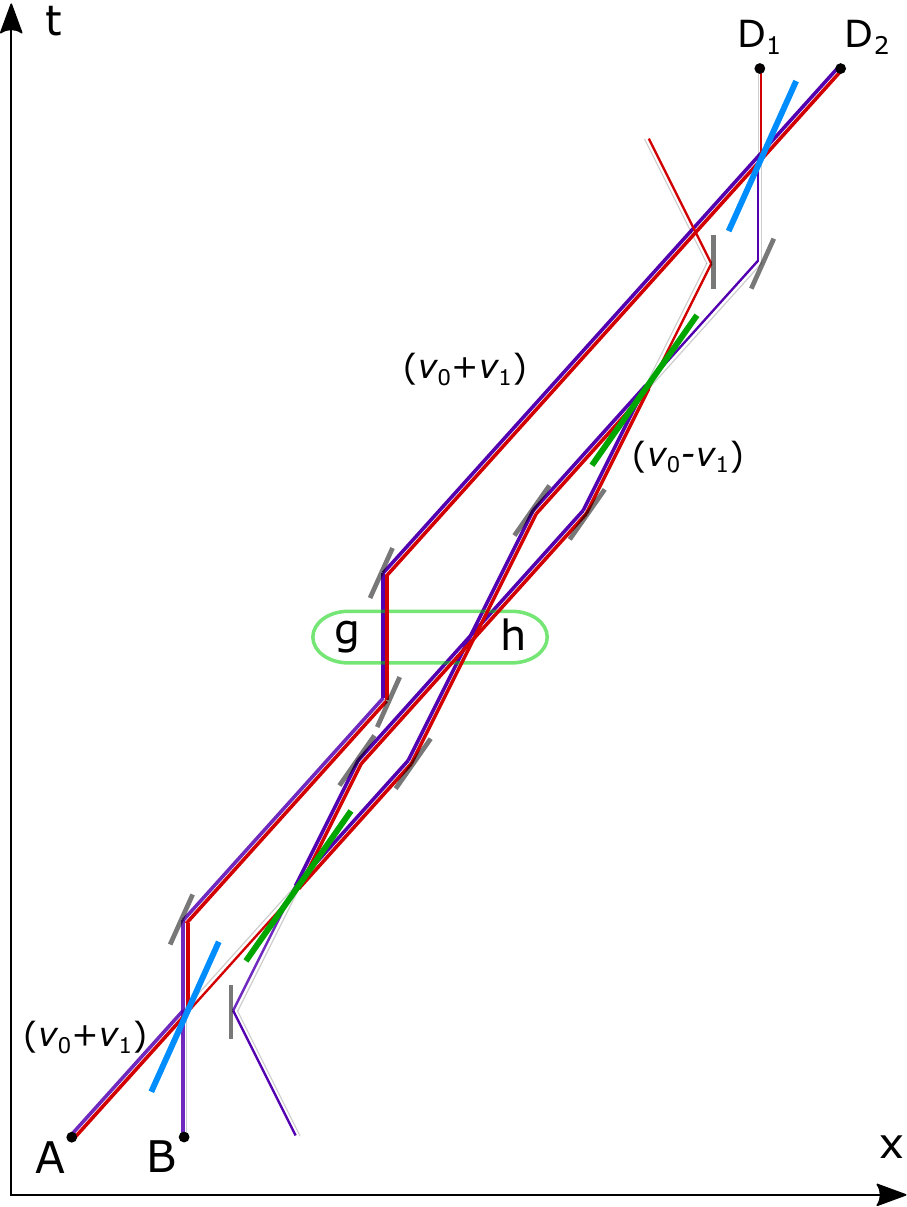}
    \caption{The experimental setup with all trajectories forward from the pre-selection in red, and all trajectories backward from the post-selection in dark purple.  Nonzero weak values are only found in the regions where the two overlaps, which are shown with thicker lines. }
    \label{PPS_PrePost}
\end{figure}

We now have all of the components of our apparatus, and we know how they will interact with incoming particles at velocities $v_+$ and $v_-$, which is everything we need for the experiment.  Consider the pre-selection where the particle is sent in along path $A$.  Fig. \ref{PPS_Pre} shows trajectories for all parts of the superposition state that propagate through the apparatus from $A$.  We can see that the upper path leads to the two detectors, while the lower path ultimately escapes the apparatus undetected.  For our post-selection, we consider the case that detector $D_2$ fires, which only appears to be possible if the particle traversed the upper path.  Fig. \ref{PPS_Post} shows trajectories for all parts of the superposition state obtained by back-propagating through the apparatus from $D_2$.  We can see that the upper path leads back to the input ports, but the lower path leads out of an unused input port of the apparatus. Fig. \ref{PPS_PrePost} shows both the pre-selected and post-selected wavefunctions.  The weak values of the projectors onto regions where the pre- and post-selection do not overlap are zero, as are the localized momentum and energy.  The weak values of these observables can be nonzero where the pre- and post-selection overlap.

To choose the reflectivity of the our first beam splitter, we need to consider the normalized pre-selection
\begin{equation}
    \psi(x) = \frac{g(x) + f_+(x) + f_-(x)}{\sqrt{3 + 2e^{-k_1^2 / 2 \Delta k_f^2}}}.
\end{equation}
The first beam splitter (speed $\frac{1}{2}v_+$) is tuned so that
\begin{equation}
    \frac{1}{3 + 2e^{-k_1^2 / 2 \Delta k_f^2}}
\end{equation}
of the intensity of the beam is reflected ($v_r=0$), and 
\begin{equation}
    \frac{2 + 2e^{-k_1^2 / 2 \Delta k_f^2}}{3 + 2e^{-k_1^2 / 2 \Delta k_f^2}}
\end{equation}
is transmitted ($v_t = v_+$).  The second beam splitter (speed $v_0$) then transforms the $v_+$ term into an equal superposition of $v_+$ and $v_-$ (with no phase difference), which gives us $\psi$ once the packets arrive in the target region.  

To choose the reflectivity of our final beam splitter, we consider the normalized post-selection
\begin{equation}
    \phi(x) = \frac{g(x) + f_+(x) - f_-(x)}{\sqrt{3 - 2e^{-k_1^2 / 2 \Delta k_f^2}}}.
\end{equation}
The final beam splitter (speed $\frac{1}{2}v_+$)  is tuned so that fraction
\begin{equation}
    \frac{1}{3 - 2e^{-k_1^2 / 2 \Delta k_f^2}}
\end{equation}
of a back-propagated beam from $D_2$ is transmitted back toward region $g$ (speed $v_+$), and fraction
\begin{equation}
    \frac{2 - 2e^{-k_1^2 / 2 \Delta k_f^2}}{3 - 2e^{-k_1^2 / 2 \Delta k_f^2}}
\end{equation}
is reflected toward region $h$ (speed 0).  The third beam splitter (speed $v_0$) then transforms the $v_+$ term into an equal superposition of $v_+$ and $v_-$ with a $\pi$ phase difference between the terms, which gives us $\phi$ once the packets are back-propagated to the target region.

\section{Analysis}  \label{sec:analysis}

In this experiment, we find that the weak value of the particle's momentum in region $h$ is $2p_1$, while the weak value of its position places it squarely in the region $g$.  Furthermore, if the particle must follow a physically consistent trajectory from the pre-selection to the post-selection, it could only have gone by the path through region $g$, and indeed we detect no mass in region $h$.  However, if we consider the joint observables of the momentum in region $g$ and the momentum in region $h$, we find that all of the $2p_1$ momentum is located in region $h$ and none in region $g$ - so the particle is in one place, and its momentum is in another.

The weak value of the particle's kinetic energy is $\frac{1}{2m}(4p_0p_1 + \Delta p_f^2)$, and if we look at the joint observables of the energy in regions $g$ and $h$, we find that the $\frac{\Delta p_f^2}{2m}$ is in region $g$ and the $\frac{2p_0p_1}{m} $ is located in region $h$.  Since we can vary $p_0$, $p_1$, and $\Delta p_f$ more or less freely, and the regions $g$ and $h$ can be made arbitrarily far apart, it is easy to arrange it so that the vast majority of the particle's kinetic energy is located in $h$ (along with its momentum), even though the particle is located in $g$.

As in Vaidman's original nested interferometer setup \cite{vaidman2013past, ben2017improved}, there are regions where the particle leaves no weak trace on the way in or out, but does leave weak traces between those region, where it should not have been able to go.  If we consider the pre- and post-selection, and the mirrors and beam splitter of the outer interferometer to be `first', we can then make a choice about whether to add the inner interferometer `second.'  If all mirrors and beam splitters of the inner interferometer are absent, only the pre-and post-selected paths through region $g$ overlap, so all weak values measured outside region $g$ would be zero all the way through, so it looks as if there is nothing there.  What is remarkable is that if our moving inner interferometer is present, we have added some measurable energy and momentum to this nothing.

To see how this works, it is also helpful to consider the weak traces left on the arms of the inner interferometer when the packets are well-separated (outside the region $h$).  
The packets $f_+$ and $f_-$ are only truly separate and orthogonal in the limit $k_1^2/\Delta k_f^2 \rightarrow \infty$, so we consider this case first.
In this limit, we can define projectors onto $f_+$ and $f_-$,
\begin{equation}
    \begin{array}{cc}
       \hat{\Pi}_{f_+} |\psi\rangle = |f_+\rangle,  & \hat{\Pi}_{f_-} |\psi\rangle = |{f_-}\rangle, \\\\
       \langle \phi |\hat{\Pi}_{f_+}^\dag  = \langle {f_+}|,  & \langle \phi | \hat{\Pi}_{f_-}^\dag  = -\langle{f_-}|,
    \end{array}
\end{equation}
and evaluate their weak values and the corresponding energy and momentum weak values,
\begin{equation}
\begin{array}{cc}
        \big(\hat{\Pi}_{f_\pm}^\dag\hat{\Pi}_{f_\pm}\big)_w = \pm1, &  \big(\hat{\Pi}_{f_\pm}^\dag p\hat{\Pi}_{f_\pm}\big)_w = \pm(p_0 \pm p_1), \nonumber
\end{array}
\end{equation}
\begin{equation}
    \frac{1}{2m}\big(\hat{\Pi}_{f_\pm}^\dag p^2\hat{\Pi}_{f_\pm}\big)_w = \pm \frac{1}{2m}\big((p_0 \pm p_1)^2 + \Delta p_f^2\big).\label{fweak}
\end{equation}.


Thus, the projectors onto the separate functions have equal and opposite weak values, just as in Vaidman's case.  At the crossing point ($h$) the positive and negative projector weak values cancel out, while the energy and momentum remain finite.

When the packets are well-separated, we have effectively three locations to find the particle, $g$, $f_+$, and $f_-$, and the projector weak values are +1, +1, and -1, respectively, so we have a case of the 3-box paradox \cite{waegell2023quantum}.


The presence of negative projector weak values also establishes that quantum contextuality \cite{KS, spekkens2005, pusey2014anomalous, kunjwal2019anomalous} plays some role in this effect, which is unsurprising given its known relation to other counterintuitive results obtained using time-symmetric quantum mechanics \cite{waegell2017confined, waegell2018contextuality}.

The original Quantum Cheshire Cat was an example of confined contextuality in the 2-qubit Peres-Mermin (PM) square \cite{waegell2017confined, waegell2018contextuality, yu2014quantum}, which is a well-known proof of the Kochen-Specker theorem.  The pre-selection is a joint eigenstate of the three observables in one row of the PM square, and thus assigns eigenvalues to all three, and the post-selection is a joint eigenstate of the observables in a second row.  Since at least one of the nine observables in the square cannot be assigned a noncontextual eigenvalue, this pre- and post-selection confines the offending observable to the third row of the square, which is evidenced by the joint eigenprojectors of the observables in that row having some anomalous weak values.  Another approach to seeing contextuality in the original experiment and its connection to weak values is given in \cite{hance2023contextuality}

In any event, attempts to assign eigenvalues to all observables fail because they lead to logical contradictions, which do not seem to correspond to any sensible state of the physical systems in question \cite{hofmann2015quantum}.  So proofs of contextuality tell us how nature is not, but they say nothing constructive about how nature actually is.  Both the original and present versions of the Quantum Cheshire Cat, as well as all related pre- and post-selection paradoxes are attempts to interpret all weak values that are validated by the ABL rule as simultaneous physical properties of the system, even though these properties are mutually contradictory - like having one particle with its momentum in one place and its mass in another.

\section{The Counterparticle Picture} \label{sec:counter}

As discussed, there are other way to interpret these weak values, and each may lend different intuition about the underlying physics.

In the retrocausal weak-value-based picture of Aharonov, et al. \cite{aharonov2018weak, waegell2023quantum}, the weak values in this experiment are explained by the presence of an extra positive-negative pair of (counter)particles that only exist during the time between the pre- and post-selection.  In this model, each particle carries all of its physically properties locally along a well-defined trajectory, and the set of all properties has a noncontextual weak value assignment.  In general, it is noncontextual eigenvalue assignments that are ruled out by the Kochen-Specker theorem, so there is no contradiction.  Here there is no use of the ABL rule, nor is there an attempt to assign eigenvalues to observable properties.  Instead the complex weak values are interpreted as the weakly observable properties during the time between pre- and post-selection, and extra positive-negative pairs of real or imaginary particles are introduced to explain what is observed.  This constructive approach resolves the logical contradictions of assigning eigenvalues, at the costs of retrocausality needed to define the weak values, and the introduction of exotic new counterparticles.

In this experiment, the extra positive-negative pair are created at source A (see Fig. \ref{PPS_PrePost}) during the pre-selection, propagate together into the outer interferometer (leaving no trace), take separate paths through the inner interferometer (crossing in region $h$), and then propagate together through the outer interferometer (again leaving no trace) and are annihilated at D$_2$ during the post-selection.

Inside the inner interferometer, the negative particle carries negative kinetic energy, $-\frac{1}{2m}\big((p_0 - p_1)^2 + \Delta p_f^2\big)$ and negative momentum (opposite the direction of propagation), $-(p_0 - p_1)$ and the positive particle carries positive kinetic energy, $\frac{1}{2m}\big((p_0 + p_1)^2 + \Delta p_f^2\big)$ and momentum, $(p_0 + p_1)$, but their magnitudes are not equal, resulting in net positive momentum and energy.  In the limit $k_1^2/\Delta k_f^2 \rightarrow \infty$, these are the weak values for packets $f_+$ and $f_-$, and they sum to the values found for region $h$ (even outside the limit).  As such, we will argue that even outside the limit we still have this positive-negative pair of counterparticles propagating through the apparatus, in addition to some other complicated effects due to the cross-terms that we leave for future analysis.

Thus both the weak values measured when the packets are well-separated and the weak values measured when they fully overlap are explained by the wave packets of a positive-negative pair of counterparticles passing through the apparatus.

A second argument for the presence of the positive-negative pair can be given by modifying the experiment to include entanglement with the internal spin of the particle.  If we choose the pre-selection and post-selection to be
\begin{equation}
    |\psi\rangle = g|\uparrow\rangle + f_+|\rightarrow\rangle + f_-|\leftarrow\rangle,
\end{equation}
and
\begin{equation}
    |\psi\rangle = g|\uparrow\rangle + f_+|\rightarrow\rangle - f_-|\leftarrow\rangle,
\end{equation}
then the orthogonal spin states allow us to separate $f_+$ and $f_-$, even outside the limit where $k_1^2/\Delta k_f^2 \rightarrow \infty$, we get all of the same weak values of Eq. \ref{fweak}, and there are no cross terms.

A positive-negative pair of particles also explains the weak values in Vaidman's original nested interferometer \cite{vaidman2013past,aharonov2018weak,waegell2023quantum}, and a pair with net negative energy and momentum explains the weak values in interaction free energy transfer \cite{elouard2020interaction,waegell2020energy}.

\section{Conclusions}  \label{sec:conc}

We have presented a variant of the Quantum Cheshire Cat effect where weak measurements of a pre- and post-selected system show that the mass of a particle is entirely separated from its momentum, and mostly separated from its energy.  We have also given a \textit{gedanken} experiment that realizes the necessary pre- and post-selection using nested interferometers which are in motion in the laboratory frame.  This is yet another counterintuitive prediction of quantum theory which is relatively straightforward in the two-vector quantum formalism of ABL, but would have been difficult to conceive in the usual one-vector quantum formalism.  This result adds to a growing list of remarkable quantum phenomena discovered in this way.

Finally, we have shown how the retrocausal counterparticle model of Aharanov et al. provides an intuitive explanation of the relevant weak values in this experiment, using particles that move on definite trajectories through the experiment.\newline

\textbf{Acknowledgments:} This project/publication was made possible through the support of Grant 63209 from the John Templeton Foundation. The opinions expressed in this publication are those of the authors and do not necessarily reflect the views of the John Templeton Foundation.

\bibliography{refs}

\end{document}